\documentstyle[12pt,epsf]{article}
\newcommand{\beq}{\begin{equation}}
\newcommand{\eeq}{\end{equation}}
\newcommand{\bea}{\begin{eqnarray}}
\newcommand{\eea}{\end{eqnarray}}

\begin{document}
{\thispagestyle{empty}
\noindent \hspace{1cm}  \hfill October 1994 \hspace{1cm}\\
\mbox{}                 \hfill BI-TP 94/54    \hspace{1cm}\\

\begin{center}\vspace*{1.0cm}
{\large\bf Exact solution (by algebraic methods) of the lattice}\\
{\large\bf Schwinger model in the strong-coupling regime}
\\\vspace*{1.0cm}
{\large F.~Karsch, E. Meggiolaro}\\
\vspace*{0.5cm}{\normalsize
{Fakult\"at f\"ur Physik, \\
Universit\"at Bielefeld, \\
D--33615 Bielefeld, Germany.}}\\
\vspace*{1.0cm} 
{\large L. Turko}\\
\vspace*{0.5cm}{\normalsize
{Institute of Theoretical Physics,\\ 
Wroclaw University,\\ 
50--204 Wroclaw, Poland.}}\\ 
\vspace*{2cm}{\large \bf Abstract}
\end{center}

\setlength{\baselineskip}{1.3\baselineskip}
Using the monomer--dimer representation of the lattice Schwinger model,
with $N_f =1$ Wilson fermions in the
strong--coupling regime ($\beta=0$), we evaluate its partition function,
$Z$, exactly on finite lattices. By studying the zeroes of $Z(k)$
in the complex plane $(Re(k),Im(k))$ for a large number of small
lattices, we find the zeroes closest to the real axis for
infinite stripes in temporal direction and spatial extent $S=2$ and 3.
We find
evidence for the existence of a critical value for the hopping parameter
in the thermodynamic limit $S\rightarrow \infty$ on the real axis
at about $k_c \simeq 0.39$.
By looking at the behaviour of quantities, such as the chiral
condensate, the chiral susceptibility and the third derivative of $Z$
with respect to $1/2k$, close to the critical point $k_c$,
we find some indications for a continuous phase transition.
}
\vfill\eject

\section{Introduction}

The Schwinger model \cite{Schwinger62}, that is 2--dimensional QED (QED$_2$)
with massless electrons, has always attracted
the interest of theoreticians not only because it is a rather simple model,
which can be solved analytically, but also (and mainly!) because many of its
properties are quite similar to those of 4--dimensional QCD.
Already in the basic version with $N_f = 1$ fermion flavours, one recovers
a lot of QCD--like properties such as confinement for fermions, chiral
symmetry breaking, due to the anomaly in the $U(1)$ axial current, and
charge screening
\cite{Schwinger62,Lowenstein-Swieca71,Casher-et-al.74}.

By virtue of this similarity, one is also tempted to consider the lattice
version of the Schwinger model as a test--model for lattice
4--dimensional QCD (QCD$_4$). Also here one faces the problem of the
choice
of a lattice scheme for fermions: the two most common choices are the
well--known {\it Wilson fermions} \cite{Wilson75} and the {\it staggered}
(or {\it Kogut--Susskind} [KS])
{\it fermions} \cite{Susskind77,Banks-et-al.77}.
Most of the lattice calculations done up to now for
the Schwinger model used the staggered fermion formulation
\cite{Marinari-et-al.81,Carson-Kenway86,Dilger92}, in which case the
chiral limit is obtained by simply setting the bare fermion mass
parameter $m$ appearing in the lattice Lagrangian to zero: $m \to 0$.
All these lattice calculations seem to reproduce well
the expected properties of the continuum massless Schwinger
model, known from analytical results.

On the contrary, very little is known about the lattice Schwinger model
with
Wilson fermions. Our interest in the Wilson formulation of QED$_2$ comes
from the recently made observation \cite{Gausterer-et-al.92} that the
critical point in the hopping parameter, at which the chiral limit is
reached, may not agree with the {\it na{\"{\i}}ve} expectation 
(see below, for a more detailed discussion of this point).  
This may also be of relevance for an
understanding of the complicated phase diagram found in the Wilson
formulation of lattice QCD$_4$ with large number of flavours
\cite{Iwasaki-et-al.92}.

In the lattice action with Wilson fermions the bare fermion mass
$m$ does not appear explicitly as in the lattice action with KS fermions.
It contains as parameters the coupling constant $\beta$ and
the {\it hopping parameter} $k$, which is related to the bare fermion mass.
Therefore in the Wilson fermion formulation there is the problem of
defining a chiral limit, corresponding to $m \to 0$.
In lattice QCD$_4$ with Wilson fermions the chiral limit, at a
given value of $\beta$, is reached when the hopping parameter $k$
approaches
a certain critical value $k_c(\beta)$, often defined as the value of $k$
for which the pion mass $M_\pi$ vanishes. In fact, in the chiral limit of
QCD$_4$ the pion becomes the Goldstone boson of the spontaneously--broken
chiral symmetry and its mass is expected to vanish as $\sqrt{m}$ when
$m \to 0$. It tacitly is assumed that this defines a critical point
which coincides with a critical point (zero) of the partition
function.

It is expected that the situation is similar for the lattice
Schwinger model, in the sense that there will be a critical point for
each value of the gauge coupling,$k = k_c(\beta)$, which defines
the chiral limit. The continuum chiral limit will be reached following
this line up to $k_c(\beta = \infty) = 1/2d = 1/4$.
However, we cannot determine $k_c(\beta)$ in the same way as in QCD$_4$,
since in the Schwinger model with $N_f = 1$ we have no Goldstone boson
in the chiral limit $m \to 0$: the $U(1)$ chiral symmetry is broken by the
anomaly and one is left with a massive pseudoscalar, similar to the
$\eta'$ in QCD$_4$. A determination of $k_c(\beta)$ thus has to proceed
through the direct investigation of the singularities of $Z(k)$.

A quite common attitude is to assume that $k_c(\beta)$ coincides with the
convergence radius $\bar{k}(\beta)$ for the joint expansion in the hopping
parameter $k$ and the inverse gauge coupling $\beta$. However, there
is no proof that this is correct: for example, in the
lattice Schwinger model with KS fermions we have that $\bar{k} \le 1/2$
\cite{Gausterer-et-al.92}, while $k_c = \infty$.
For Wilson fermions, one does not know the precise values of
$\bar{k}(\beta)$ and $k_c(\beta)$.
Our aim is to compare these two values at least in the strong coupling
regime $\beta = 0$. It was already found in ref. \cite{Gausterer-et-al.92}
that $\bar{k}(0) \le 1/2$ and some indications from Monte Carlo simulations
were reported, indicating that $k_c(0) \ne \bar{k}(0)$.

In this work we will determine $k_c(0)$ by deriving analytically (using
algebraic methods) the Lee--Yang zeroes \cite{Lee-Yang52} of the partition
function $Z(k)$ for the lattice Schwinger model, with $N_f = 1$
Wilson fermions, in the strong--coupling regime ($\beta = 0$).
For a finite lattice $S \times T$ these
zeroes have a non--vanishing imaginary part, in the complex plane
$(Re(k),Im(k))$, indicating that there is no
real critical point for a finite lattice. This remains true for
$T\rightarrow \infty$ with finite $S$.
Yet, enlarging the lattice, they
show a tendency to move towards the real $k$--axis.
By studying the zeroes in the complex plane of the
partition function $Z(k)$ for a large number of small lattices, and then
extrapolating to the thermodynamic limit $\infty \times \infty$, we will
find evidence for the existence of a real critical value for the hopping
parameter, at about $k_c \simeq 0.39$.
We will also study some relevant quantities, such as the chiral
condensate $<\bar{\psi}\psi >$, the chiral susceptibility and the third
derivative, with respect to $1/2k$, of the partition function,
in order to get some information about the question of the
order of the phase transition.

\section{The method}

The action for the lattice Schwinger model is written as the sum of a
gauge--action $S_G[U]$ and of a fermion--action $S_F[\psi,\bar{\psi},U]$:
\beq
S = S_G[U] + S_F[\psi,\bar{\psi},U].
\eeq
The gauge part $S_G[U]$ is given by:
\beq
S_G[U] = \beta \displaystyle\sum_{P}[1-{1 \over 2}(U_P + U^{\dagger}_P)],
\eeq
where $\beta \equiv 1/e^2$, $e$ being the usual electro--magnetic coupling
constant. Since the gauge group is $U(1)$, the basic lattice gauge variable
$U_\mu (n)$, corresponding to the link connecting the sites $n$ and $n +
{\mathaccent 94 \mu}$, can be written in the form of a phase:
\beq
U_\mu (n) = \exp[i\phi_\mu (n)].
\eeq
In eq.(2) $U_P$ stands for the usual $1 \times 1$ Wilson plaquette,
constructed using the link variables $U_\mu (n)$.

In the strong coupling limit ($e^2 \to \infty$), the coefficient $\beta$
in front of $S_G[U]$ goes to zero and the total action $S$ reduces simply
to the fermion action $S_F[\psi,\bar{\psi},U]$.
The action $S_F$ for Wilson fermions (and only one flavour) can be
written in the form:
\beq
S_F = {1 \over 2k}\displaystyle\sum_{n,m}\bar{\psi}(n)
K_{nm}[U]\psi (m) ,
\eeq
where $k$ is the so--called {\it hopping parameter}, and each $K_{nm}$,
for a given couple of lattice sites $n$ and $m$, is a matrix in Dirac space:
\bea
\lefteqn{
K_{nm}[U] = } \nonumber \\
& & \delta_{nm} \cdot {\bf I} -k\displaystyle\sum_{\mu} [(r-\gamma_\mu)
U_\mu (n) \delta_{n+{\mathaccent 94 \mu},m} +
(r+\gamma_\mu)
U^\dagger_\mu (n-{\mathaccent 94 \mu}) \delta_{n-{\mathaccent 94 \mu},m}]
.
\eea
Therefore $K_{nm}$ is of the form:
\beq
K_{nm} = \delta_{nm} \cdot {\bf I} - k \cdot M_{nm}[U] ,
\eeq
where the only non--vanishing matrices $M_{nm}$ are those connecting
neighbouring lattice sites:
\bea
M_{n,n+{\mathaccent 94 \mu}}[U] &=& (r-\gamma_\mu)U_\mu (n) ,\nonumber\\
M_{n,n-{\mathaccent 94 \mu}}[U] &=& (r+\gamma_\mu)U^\dagger_\mu
(n-{\mathaccent 94 \mu}) .
\eea
The parameter $r$, which satisfies $|r| \le 1$, is called the {\it Wilson
parameter}.
In the following we will consider only the case $r = 1$.
The matrices $\gamma_\mu$, with $\mu = 1,2$ are the $2 \times 2$
Euclidean Dirac matrices, corresponding to $1 + 1$ space--time dimensions
(in particular we will consider the index $\mu = 1$ as corresponding to the 
time dimension and the index $\mu = 2$ as corresponding to the space 
dimension): they satisfy the anticommutation relation 
$\{ \gamma_i ,\gamma_k \} = 2 \delta_{ik} {\bf I}_2$, with ${\bf I}_2$ 
being the $2 \times 2$ identity matrix.
For our algebraic manipulations, we have chosen the following
representation for the $\gamma$--matrices:
\beq
\gamma_1 = \pmatrix{1 & ~0 \cr 0 & -1 \cr} ~~~ , ~~~
\gamma_2 = \pmatrix{0 & 1 \cr 1 & 0 \cr} .
\eeq
Finally, the partition function, in the strong coupling regime $\beta = 0$,
is given by the following expression:
\bea
Z(k) &=& \int [D\bar{\psi} D\psi ] \int [DU] e^{-S_F[\psi,\bar{\psi},U]}
\nonumber \\
 &=& \int [D\bar{\psi} \psi ] \int [DU] e^{-{1 \over 2k}\displaystyle\sum_{n,m}
\bar{\psi}(n) K_{nm}[U]\psi (m)} .
\eea
Following the standard normalization convention, we also eliminate the 
factor $1/2k$ appearing in the exponent in eq.(9) by re--scaling the fermion
fields with $\sqrt{2k}$:
\beq
\psi = \sqrt{2k}\tilde{ \psi},~~~\bar{\psi} = \sqrt{2k}
\bar{\tilde{ \psi}} .
\eeq
When evaluating a matrix element of the form
$<\prod_{n=1}^{N}\tilde{\psi}_{A_n}(x_n) \bar{\tilde{\psi}}_{B_n}(y_n)>$,
in terms of the re--scaled fields, we must however keep in mind that the 
original correlation function is obtained by multiplying with $(1/2k)^N$.
When considering a lattice with $S$ lattice sites in the space direction
and $T$ lattice sites in the time direction, for a total of
$N = S \times T$ lattice sites, the partition function (9) becomes:
\beq
Z(k) = (2k)^{-2N} \tilde{ Z}(k) ,
\eeq
where we have defined:
\beq
\tilde{ Z}(k) \equiv
\int [D\bar{\tilde{ \psi}}
D\tilde{ \psi} ] \int [DU]
e^{-\displaystyle\sum_{n,m} \bar{\tilde{ \psi}}(n) K_{nm}[U]
\tilde{ \psi} (m)} .
\eeq
It is exactly $\tilde{ Z}(k)$ that we have evaluated with algebraic
methods for
a large series of small lattices. First of all we have put
$\tilde{ Z}(k)$ in a
comfortable form for subsequent algebraic manipulations.
Making use of the explicit expression (3) for the link variables $U$ and
remembering the Grassmann properties of the fermion fields, one finds
the monomer--dimer representation \cite{Rossi-Wolff84}
for the partition function $\tilde{ Z}(k)$:
\beq
\tilde{ Z}(k) \equiv  \int [D\bar{\tilde{ \psi}}
D\tilde{ \psi} ]
\displaystyle\prod_{n} F(n) L_1(n) L_2(n) ,
\eeq
where $F(n)$ is the monomer term at lattice site n, coming from the
mass--term in the action ({\it
i.e.} the bilinear diagonal term not containing the gauge fields $U$):
\beq
F(n) = 1 - \bar{\tilde{ \psi}}_1 (n) \tilde{ \psi}_1 (n) -
\bar{\tilde{ \psi}}_2 (n) \tilde{ \psi}_2 (n)
 + \bar{\tilde{ \psi}}_1 (n) \tilde{ \psi}_1 (n)
\bar{\tilde{ \psi}}_2 (n) \tilde{ \psi}_2 (n) .
\eeq
(The indices $1$ and $2$ are the Dirac indices). The quantities
$L_\mu (n)$, with $\mu = 1,2$, are the dimer terms defined on links of
the lattice. They result from the direct integration over the
gauge field $U_\mu (n) = exp[i\phi_\mu (n)]$, corresponding to the link
$n \to n + {\mathaccent 94 \mu}$. Explicitly:
\beq
L_\mu (n) = \displaystyle\int_{-\pi}^{\pi} {d\phi_\mu (n) \over 2\pi}
e^{\{ M_\mu (n) exp[i\phi_\mu (n)] + N_\mu (n) exp[-i\phi_\mu (n)]\} } ,
\eeq
where $M_\mu (n)$ and $N_\mu (n)$ are given by:
\bea
M_\mu (n) &=& k \bar{\tilde{ \psi}}(n)(1-\gamma_\mu)
\tilde{ \psi}(n+{\mathaccent 94 \mu}) , \nonumber \\
N_\mu (n) &=& k \bar{\tilde{ \psi}}(n+{\mathaccent 94 \mu})
(1+\gamma_\mu)\tilde{ \psi}(n) .
\eea
Thanks to the particularly simple form (3) of the gauge variables $U$
in the case of a $U(1)$ gauge group, the integration in (15) can be 
performed in an elementary way, after having expanded as a power--series 
the first exponential.
Making use of the explicit representation (8) for the $\gamma$--matrices
and of the Grassmann properties of the fermion fields, one finds the
following rather simple expressions for the one--link integrals $L_1(n)$
and $L_2(n)$:
\bea
L_1 (n) &=& 1 + M_1(n) N_1(n) \nonumber \\
    &=& 1 + 4 k^2 {\bar{\tilde{ \psi}}}_2(n) \tilde{ \psi}_2
(n + {\mathaccent 94 1})
                 {\bar{\tilde{ \psi}}}_1(n + {\mathaccent 94 1})
\tilde{ \psi}_1 (n) ,
\nonumber \\
L_2 (n) &=& 1 + M_2(n) N_2(n) \nonumber \\
&=& 1 + k^2 \bar{\tilde{ \psi}}(n) \pmatrix{~1 & -1 \cr -1 & ~1 \cr}
\tilde{ \psi} (n + {\mathaccent 94 2}) \cdot
\bar{\tilde{ \psi}}(n + {\mathaccent 94 2}) \pmatrix{1 & 1 \cr 1 & 1
\cr}
\tilde{ \psi} (n) .
\eea
After inserting the expressions (14) and (17)
in (13), one is left with an integral over the fermionic
variables, which must be evaluated according to the integration rules
for the Grassmann variables. Doing this ``manually'' turns out to be
extremely boring and time--consuming, even for the very small $2 \times 2$
lattice. For this reason we have developed an algebraic method for
evaluating the integral (13). We have used the algebraic computer language
``{\it Mathematica}'' and have implemented the basic rules for the
Grassmann algebra:
\beq
\{ \eta_i, \eta_j \} = 0, ~~ \int d\eta_i = 0, ~~
\int d\eta_i \eta_k = \delta_{ik} .
\eeq
In this way we were able to write computer programs (for ``{\it
Mathematica}'') for evaluating products of polynomials of arbitrary
strings of Grassmann variables (like those appearing in eq.(13)) and for
integrating them. In practice we have used the following strategy for
calculating the partition function (13) for a given lattice of size 
$S \times T$.
First of all we have computed the {\it transfer matrix, i.e.}
the products of all one--link--terms, $L_1(n)$
and $L_2(n)$, and all mass--terms $F(n)$ belonging to a given space--like
line $(x,t)$ with $x=1,2, \ldots ,S$. We  call this object a ``{\it
line}''. It is a function of all Grassmann fields belonging to the sites
of the line under consideration:
\beq
line[t,0] = \displaystyle\prod_{x=1}^{S} F(x,t) L_1(x,t) L_2(x,t) .
\eeq
In evaluating this product we have already taken into account the {\it
torus--like} topology of the lattice, in form of periodic boundary
conditions for the Grassmann fields along space--like lines.
The partition function is represented in terms of $line[t,0]$ as:
\beq
\tilde{ Z}(k) =  \int [D\bar{\tilde{ \psi}}
D\tilde{ \psi} ]
\displaystyle\prod_{t=1}^{T} line[t,0] .
\eeq
Starting with the object $line[t,0]$ one can now
evaluate composite objects, such as the product of two adjacent
lines: $line[t,0] \cdot line[t+1,0]$. By virtue of eqs.(19) and (17),
all other lines, different from $line[t,0]$ or $line[t+1,0]$, do not
depend on the Grassmann
fields on the sites $(x,t+1), ~~x=1,2, \ldots ,S$. Therefore we can
integrate the product $line[t,0] \cdot line[t+1,0]$ with respect to
these Grassmann fields and obtain:
\beq
line[t,1] = \int \displaystyle\prod_{x=1}^{S} d\bar{\tilde{
\psi}}(x,t+1) d\tilde{ \psi}(x,t+1) ]  line[t,0] \cdot line[t+1,0] .
\eeq
We can then proceed in the same way and construct more extended objects.
For example we can multiply $line[t,1]$ with $line[t+2,0]$, or even with
$line[t+2,1]$, and integrate over the Grassmann fields lying on the line
$(x,t+2), ~~x=1,2, \ldots ,S$: In general we obtain $line[t,l]$ by
performing the integration over the ($l$) lines of
intermediate Grassmann fields at $x=t+1,...,t+l$.
Finally, for $l=T-1$ the resulting object covers just the entire
lattice $S \times T$ and we must again take into account the {\it
torus--like} topology of the lattice, {\it i.e.} impose anti--periodic
boundary conditions for
the Grassmann fields along time--like lines (In fact, it turns out to be
irrelevant, for the final result, if we impose periodic or anti--periodic
boundary conditions for these Grassmann fields). In the last step, we can
integrate over the remaining Grassmann fields and obtain the final
result for the partition function (13).

It turns out that the CPU time required in this approach is entirely
controlled by the initial spatial extent $S$ of the lattice as this
determines the number of Grassmann fields one can combine in a given
string of fields. Doubling the length in time direction does not lead to
a drastic increase of the computer time as the resulting object,
$line[t,2l]$ contains exactly the same number and types of strings
of Grassmann fields as $line[t,l]$: the only additional complication
results from the more complex structure of the coefficients of these
strings, which become higher order polynomials in $k$ for increasing $T$.
As a curiosity, if we try to follow the strategy of ``{\it extending}'' in
the space--direction, instead of extending in the time--direction, we need
(for our programs) CPU times which are orders of magnitude larger than
in
the previous case. This is simply due to the space--time asymmetry in the
representation (8): but of course the partition function for a lattice
$S \times T$, namely $\tilde{Z}(k,S,T)$, is exactly equal to the partition
function $\tilde{Z}(k,T,S)$ for a lattice $T \times S$ (One can always choose
a representation for the $\gamma$ matrices in which $\tilde{\gamma}_1 =
\gamma_2$ and $\tilde{\gamma}_2 = \gamma_1$).

\section{Results}

Following the computational strategy discussed in the previous section,
we have evaluated the partition function $\tilde{ Z}(k)$, given by (13),
for a large number of lattices of the form $S \times T$ with $S=2$ and
3 ranging from $2 \times 2$ up to $2 \times 32$ and from
from $3 \times 3$ up to $3 \times 16$. These calculations could be
performed on a workstation. For $S\ge 4$ considerably more computer
time and memory would be required.

From the Grassmann properties of the fermion fields, it immediately follows
that the function $\tilde{ Z}(k,S,T)$, for a given lattice having
$N = S \times T$
sites, is a polynomial of order $2N$ in the hopping parameter $k$,
\beq
\tilde{ Z}(k,S,T)=
\displaystyle\sum_{n=0}^{N} a_{2n} k^{2n} ,
\eeq
On a lattice with $N$ lattice sites there are
$4N$ different fermion fields (four fields for each site), so that, by
virtue of the properties of the Grassmann algebra, one can construct strings
of fermion fields with at most $4N$ fields. And since each power $k^2$ is
always accompanied by four fermion fields, one can at most construct a
string with $4N$ fields with a coefficient (proportional to ) $k^{2N}$
in front of it.

We have computed the partition functions $\tilde{ Z} (k,S,T)$
for $(S=2,T=2, \ldots ,32)$ and for $(S=3,T=3, \ldots ,16)$: each of them
is a polynomial of order $2N = 2(S \times T)$ in $k$, with  $a_0=1$
($\tilde{ Z}(k=0,S,T)=1$) and $a_{2n} \ge 0$ for all $n$.
The magnitude of these coefficients generally
increases with the order of $k$, apart from deviations in the
very first coefficients due the {\it torus--like} topology of the lattice
which allows for special ``trajectories'', made up
of chains of links wrapping around the lattice.
In Table 1 and Table 2 we report the list of these coefficients for the
two lattices $2 \times 32$ and $3 \times 16$ respectively. Note that for
$S$ and $T$ even
the partition function $\tilde{Z}(k,S,T)$ is a polynomial of
the form $\sum_{n=0}^{N/2} a_{4n} k^{4n}$. This is due to the larger
set of symmetries on such lattices, as will be discussed below.
Let us first discuss the distribution of zeroes of the partition function.
In [Fig.1] and in [Fig.2] we show  the distribution of the complex zeroes,
$(Re(k),Im(k))$, of the partition function
$\tilde{ Z}(k,S,T)$ 
for the $2 \times 32$ and $3 \times 16$ lattices, respectively.

By virtue of (22), $\tilde{Z}(k)$ is a polynomial in $k^2$ with real 
coefficients: so, if $\bar{k}$ is a complex zero of $\tilde{Z}(k)$, also
$-\bar{k}$ and $\bar{k}^*$ (the complex--conjugate of $\bar{k}$) will be
zeroes of $\tilde{Z}(k)$. As a consequence of this, the distribution of 
zeroes $(x,y)$ in the complex plane $(Re(k),Im(k))$ is invariant under the
{\it parity} transformation $(x,y) \to (-x,-y)$ ({\it P--symmetry}) and
under the {\it complex--conjugate} transformation $(x,y) \to (x,-y)$
({\it C--symmetry}). In other words, the distribution of zeroes turns out 
to be symmetric under reflections with respect to the real and/or the 
imaginary $k$--axis: this is evident from [Fig.1] and [Fig.2].
The distribution of zeroes in [Fig.1], for the lattice $2 \times 32$, has 
an additional symmetry under reflections with respect to the axis
$Re(k)-Im(k)=0$ and/or to the axis $Re(k)+Im(k)=0$.
This additional symmetry is typical for lattices of size $S \times T$, 
where both $S$ and $T$ are even numbers. In fact, 
it turns out that for this class of lattices the partition function
$\tilde{Z}(k)$ may be written in the following form:
\beq
\tilde{ Z}(k) =  \int \displaystyle\prod_{l~odd}
d\bar{\tilde{ \psi}}(l) d\tilde{ \psi}(l)
\displaystyle\prod_{m~odd} F(m)
\displaystyle\prod_{n~even} cross(n) ,
\eeq
where a given site $(s,t)$ (with $s=1, \ldots ,S$ and $t=1, \ldots ,T$) is 
said to be {\it even} or {\it odd} if the integer number $s+t$ is
respectively even or odd. While $F(m)$ is the usual monomer term which we 
have introduced before in eq.(14), $cross(n)$ is a new object obtained by 
multiplying the monomer term in the site $n$ with the four dimer terms
starting from or ending at the site $n$, and finally integrating with 
respect to the fermion fields in $n$:
\beq
cross(n) = \int d\bar{\tilde{ \psi}}(n) d\tilde{ \psi}(n)
F(n) L_1(n) L_2(n) L_1(n - {\mathaccent 94 1})
L_2(n - {\mathaccent 94 2}) .
\eeq
By explicitely evaluating this expression, one finds that $cross(n)$
may be written as $1 + k^4 \alpha(\bar{\tilde{ \psi}} \tilde{ \psi})$, 
where $\alpha(\bar{\tilde{ \psi}} \tilde{ \psi})$ is a sum of products of 
four fermion fields defined in the neighbouring sites of $n$ ({\it i.e.},
$n + {\mathaccent 94 1}$, $n - {\mathaccent 94 1}$, $n + {\mathaccent 94 
2}$ and $n - {\mathaccent 94 2}$).
Therefore the partition function $\tilde{Z}(k)$ will be a polynomial in
$k^4$, with real (and positive) coefficients. As a consequence, if 
$\bar{k}$ is a solution of $\tilde{Z}(\bar{k})=0$, also $i\bar{k}$ will be  
a solution. In other words, the distribution of zeroes of $\tilde{Z}(k)$
in the complex $k$--plane will be invariant under the transformation
$(x,y) \to (-y,x)$: we will call this an {\it I--symmetry}.
After combining this {\it I--symmetry} with the other $P$-- and 
$C$--symmetries, one immediately recognizes that the distribution of zeroes
is invariant under reflections not only with respect to the real or the 
imaginary $k$--axes, but also with respect to the axes
$Re(k)-Im(k)=0$ and  $Re(k)+Im(k)=0$: this is in fact what one observes in 
[Fig.1].

With increasing lattice size the zero with the smallest imaginary part comes
closer to the real axis. This zero of the partition function has been
plotted in [Fig.3] for various lattices of size $S \times T$. As can be seen,
for increasing $T$ and fixed $S$ the imaginary part of the zeroes approaches
a finite, non--zero value, indicating that there is no phase transition on
infinite stripes of extent $S < \infty$.
We have then used a polynomial fit to derive, from the two previous
series of points, the asymptotic values of the roots nearest to the real
$k$--axis corresponding to the lattices $2 \times \infty$ and $3 \times
\infty$. In practice, we have performed separate fits of real
and imaginariy parts of the sequence of roots for given $S$, using the
polynomial ansatz,  $P(T) = a_0 + a_1/T + a_2/T^2 + \ldots$.
The number of coefficients in the
polynomials was chosen so as to minimize the {\it chi--squared} value.
It turns out that for the lattices with $S=2$ good fits are obtained if one 
uses a polynomial with six coefficients for the real part of the roots and a 
polynomial with five coefficients for the imaginary part. Instead, for the 
lattices with $S=3$ good fits are obtained using polynomials with only three 
coefficients for both the real and the imaginary part of the roots.

In this way we have found that the root nearest to the real $k$--axis is
approximately $(0.464 \pm 0.003, ~0.172 \pm 0.004)$ for the 
$2 \times \infty$ lattice and
approximately $(0.43 \pm 0.01, ~0.09 \pm 0.01)$ for the  
$3 \times \infty$ lattice.

If one tries to linearly extrapolate these two points, assuming the
existence of a real critical value $k_c$ in the
thermodynamic limit $S \to \infty$ and $T \to \infty$ (see also below), 
one finds approximately (within the accuracy of our linear extrapolation):
\beq
k_c \simeq 0.39~~.
\eeq
Although the spatial extent of our lattices is still quite small, it is 
interesting to analyze the complex zeroes of infinite stripes under the 
assumption of a second order phase transition in the thermodynamical limit
$S \to \infty$. The real and the imaginary parts of the zero closest to the 
real axis are then expected to scale like:
\bea
Re(k(S)) &=& k_c + {a \over S^{1/\nu}} , \nonumber \\
Im(k(S)) &=& {b \over S^{1/\nu}} .
\eea
Using this {\it ansatz} for our $S=2$ and $S=3$ results  
we find from the $S$--dependence of $Im(k)$ the critical exponent
$\nu \simeq 0.63$ and using this to extract $k_c$ from $Re(k)$ we again
find $k_c \simeq 0.39$. Of course, in order to get a realistic estimate of the
errors on these numbers, one needs to consider also lattices with much
larger $S$.

This is surely an extremely interesting result: the value
we have found for $k_c$ is considerably smaller than the value $\bar{k} =
1/2$ we would have expected from {\it na{\"{\i}}ve} argumentations.
Since for $\beta = \infty$ (theory with free fermions, since $e^2 = 1/\beta
\to 0$) it is well known that $k_c = 1/4$ (in general it will be $k_c =
1/2d$, with $d$ space--time dimensions), we are led by the result (25) to
believe in the existence of a line of phase transition from $(\beta =0,
k_c \simeq 0.39)$ to $(\beta = \infty ,k_c = 1/4)$.
(See also ref. \cite{Gausterer-Lang94}, where this same result was found
making use of the so--called {\it eight--vertex model} \cite{Salmhofer91}
for the analytical computations.)

In order to get some further information about the question of the order of 
the phase transition, we have studied the behaviour 
of the (re--scaled) chiral condensate
$<\bar{\tilde{\psi}} \tilde{\psi} >$, the (re--scaled) chiral susceptibility
$\tilde{\chi}$ and the (re--scaled) third derivative, with
respect to $1/2k$, of the logarithm of the partition function $Z$ in the
vicinity of the critical point $k_c$.
The re--scaled chiral condensate $<\bar{\tilde{\psi}} \tilde{\psi} >$ is
given by 
the following expression in terms of the partition function $\tilde{Z}(k)$
(see eqs.(5), (9) and (11)):
\beq
<\bar{\tilde{\psi}} \tilde{\psi} > =
-{1 \over (2k)N}{d \over d(1/2k)}ln(Z) =
-2 + {k \over N} \left( {1 \over \tilde{Z}(k)}{d\tilde{Z}(k) \over dk}
\right) ,
\eeq
where $N = S \times T$ is, as usual, the number of sites of the lattice.
In [Fig.4] we report the behaviour of 
$\tilde{Q} \equiv -<\bar{\tilde{\psi}} \tilde{\psi} >$
as a function of $k$, for the two lattices $2 \times 32$ and $3 \times 16$.
The re--scaled chiral susceptibility $\tilde{\chi}$ is defined as:
\bea
\tilde{\chi} &\equiv& {1 \over (2k)^2 N}{d^2 \over d(1/2k)^2} ln(Z) 
\nonumber \\
&=& \displaystyle\sum_{n} < \bar{\tilde{\psi}}(n) \tilde{\psi}(n)
\bar{\tilde{\psi}}(0) \tilde{\psi}(0) >
- N < \bar{\tilde{\psi}} \tilde{\psi} >^2 \nonumber \\
&=& -2 + {k \over N}\left[
k \left( {1 \over \tilde{Z}}{d^2 \tilde{Z} \over dk^2} \right)
-k \left( {1 \over \tilde{Z}}{d \tilde{Z} \over dk} \right)^2
+2 \left( {1 \over \tilde{Z}}{d \tilde{Z} \over dk} \right)
\right] .
\eea
In [Fig.5] we report the behaviour of $\tilde{\chi}$ as a function of $k$,
for the same two lattices. It is easy to verify that
$\tilde{\chi} = -k(d\tilde{Q}/dk) - \tilde{Q}$, so that $\tilde{\chi}$
behaves as $\tilde{\chi} \simeq -\tilde{Q} \rightarrow -2$ for $k \to 0$.
Finally, we also consider the third derivative, with respect to $1/2k$, of
the logarithm of the partition function $Z$:
\bea
\tilde{\varphi} &\equiv& -{1 \over (2k)^3 N}{d^3 \over d(1/2k)^3} ln(Z) 
\nonumber \\
&=& \displaystyle\sum_{m,n} < \bar{\tilde{\psi}}(m) \tilde{\psi}(m)
\bar{\tilde{\psi}}(n) \tilde{\psi}(n)
\bar{\tilde{\psi}}(0) \tilde{\psi}(0) >
-3N < \bar{\tilde{\psi}} \tilde{\psi} > \tilde{\chi}
- N^2 < \bar{\tilde{\psi}} \tilde{\psi} >^3 \nonumber \\
&=& -4 + {k \over N}\left[
k^2 \left( {1 \over \tilde{Z}}{d^3 \tilde{Z} \over dk^3} \right)
+6k \left( {1 \over \tilde{Z}}{d^2 \tilde{Z} \over dk^2} \right)
-3k^2 \left( {1 \over \tilde{Z}}{d \tilde{Z} \over dk} \right)
\left( {1 \over \tilde{Z}}{d^2 \tilde{Z} \over dk^2} \right) \right]
\nonumber \\
&+& {k \over N}\left[
2k^2 \left( {1 \over \tilde{Z}}{d \tilde{Z} \over dk} \right)^3
-6k \left( {1 \over \tilde{Z}}{d \tilde{Z} \over dk} \right)^2
+6 \left( {1 \over \tilde{Z}}{d \tilde{Z} \over dk} \right)
\right] .
\eea
The behaviour of $\tilde{\varphi}$ as a function of $k$ is shown in
[Fig.6], for the two lattices $2 \times 32$ and $3 \times 16$.
It is easy to demonstrate that 
$\tilde{\varphi} = k(d\tilde{\chi}/dk) + 2\tilde{\chi}$: from this one can 
immediately derive that $\tilde{\varphi}$ behaves as 
$\tilde{\varphi} \simeq 2\tilde{\chi} \rightarrow -4$ for $k \to 0$.

Clearly the three quantities are closely related. The sharpening of the
crossover in \hfill\break
$<\bar{\tilde{\psi}} \tilde{\psi} >$ is reflected in the rise of
the peak in $\tilde{\chi}$ with increasing $S$, and the narrowing of the
peak in $\tilde{\chi}$ is expressed in terms of the rapidly rising peaks
with opposite signature in $\tilde{\varphi} $. Certainly the behaviour 
of these three quantities is consistent with that expected for a continous
phase transition, {\it i.e.} a second-- or third--order phase transition.

\section{Conclusions}

We have evaluated analytically, using algebraic methods,
the partition function $Z$ for the lattice Schwinger model, with $N_f = 1$
Wilson fermions, in the strong--coupling regime ($\beta = 0$).
For a given lattice $S \times T$, the partition function is of the form
$Z(k,S,T) = (2k)^{-2N} \tilde{ Z}(k,S,T)$, where $N = S \times T$ is
the total
number of lattice sites and $\tilde{ Z}(k,S,T)$ is a polynomial in
$k$ of order
$O(2N)$. By studying the zeroes in the complex plane $(Re(k),Im(k))$ of the
partition function $\tilde{ Z}(k,S,T)$ for a large series of small
lattices
$S \times T$, we have found evidence for the existence of a critical
value for the hopping parameter, which in the thermodynamic limit,
$S,T\rightarrow \infty$ lies on the real axis at about $k_c \simeq 0.39$.
We are led by this result to believe in the existence of a line of phase
transition from $(\beta =0,k_c \simeq 0.39)$ to
$(\beta = \infty ,k_c = 1/4)$. In order to determine the order of the
transition, it is clearly important to study in more detail the density of 
the zeroes near $k_c$. This requires larger values of $S$.
We have analyzed the chiral condensate $<\bar{\psi}\psi >$, the chiral
susceptibility and the third derivative, with respect to $1/2k$, of the
partition function, in order to get some insights into the question of the
order of the phase transition. Even though the present analysis does not yet
allow to draw a definite conclusion on the order of the transition
we have found some indications that the phase transition might
be third order or even second order.

\bigskip
\noindent {\bf Acknowledgements}
\smallskip

One of us (F. K.) would like to thank C. Lang for useful discussions.
After this work was completed, we received the 
recent work by H. Gausterer and C. Lang \cite{Gausterer-Lang94},
in which they essentially find the same results as ours for $k_c$,
by using the so--called {\it eight--vertex model} for analytical computations.
This work has been partially supported by the Stabsabteilung Internationale
Beziehungen, KFA Karlsruhe, through contract number X081.29 and the EC through
contract number ERB--CHRX--CT--92--0051.

\vfill\eject

\vfill\eject
\pagestyle{empty}

\begin{table}[hbt]
\centering
\small
\setlength{\tabcolsep}{1.5pc}
\caption{The coefficients of the polynomial
$\tilde{Z}(k,2,32) = \sum_{n=0}^{32} a_{4n}k^{4n}$.}
\vspace{0.3cm}
\label{tab:table1}
\begin{tabular}{rr}
\hline
$n$ & $a_{4n}$ (for the $2\times 32$ lattice) \\
& \\
\hline
0 &  1 \\
1 &  512 \\
2 &  129024 \\
3 &  21331968 \\
4 &  2602369024 \\
5 &  249774997504 \\
6 &  19638234644480 \\
7 &  1300255043747840 \\
8 &  73961277759684608 \\
9 &  3668969473236271104 \\
10 & 160569828529865228288 \\
11 & 6255905737448198504448 \\
12 & 218535073687128684625920 \\
13 & 6883374618372455350140928 \\
14 & 196354147234439285612478464 \\
15 & 5089687819224732967794376704 \\
16 & 120172957550769852891363540992 \\
17 & 2588628553768726317536821379072 \\
18 & 50910684319442948897317601673216 \\
19 & 914109924603780980467032270045184 \\
20 & 14970604113895231454059152342515712 \\
21 & 223213979431827944404264819095502848 \\
22 & 3021076338904294430787178751238078464 \\
23 & 36958171640461419289379070931541950464 \\
24 & 406259327271308991154552486860991496192 \\
25 & 3980443123196403924685803149026276671488 \\
26 & 34377853959664866223457481686909255155712 \\
27 & 257704138028993570924127408162720968605696 \\
28 & 1639689392947634033166698976649965853474816 \\
29 & 8559749426407522788060672587795634988253184 \\
30 & 34663799157267802700683573313175561579790336 \\
31 & 97874256443064108849693716776331356231696384 \\
32 & 146811384664566690398085268830912235998019584 \\
\hline
\end{tabular}
\end{table}

\vfill\eject
\thispagestyle{empty}

\begin{table}[hbt]
\centering
\small
\setlength{\tabcolsep}{1.5pc}
\caption{The (non--zero) coefficients of the polynomial
$\tilde{Z}(k,3,16) = \sum_{n=0}^{48} a_{2n}k^{2n}$.}
\vspace{0.3cm}
\label{tab:table2}
\begin{tabular}{rr}
\hline
$n$ & $a_{2n}$ (for the $3\times 16$ lattice) \\
& \\
\hline
0 &  1 \\
3 &  1024 \\
4 &  768 \\
5 &  6144 \\
6 &  516096 \\
7 &  811008 \\
8 &  6334464 \\
9 &  174620672 \\
10 & 431947776 \\
11 & 3203923968 \\
12 & 45588021248 \\
13 & 153911033856 \\
14 & 1063856898048 \\
15 & 9906042175488 \\
16 & 40917798223872 \\
17 & 261009935695872 \\
18 & 1857033179496448 \\
19 & 8585692506488832 \\
20 & 50608039715143680 \\
21 & 303834862874263552 \\
22 & 1470365700697620480 \\
23 & 8023810955173429248 \\
24 & 42904506735217606656 \\
25 & 205822945618110185472 \\
26 & 1036084831241631694848 \\
27 & 5045499989985857634304 \\
28 & 23116972970746438483968 \\
29 & 106903512053003436687360 \\
30 & 476889028736929390133248 \\
31 & 2038448818744602390429696 \\
32 & 8604050273287356083601408 \\
33 & 34923088125822911234703360 \\
34 & 136172615876993812147470336 \\
35 & 515227117640733139176259584 \\
36 & 1857978729092258641800593408 \\
37 & 6360979177485248012381847552 \\
38 & 20656762024719021784826904576 \\
39 & 62433587250829914862254030848 \\
40 & 174406412859094327618009300992 \\
41 & 446095973409032980536257150976 \\
42 & 1025256020010205168904800567296 \\
43 & 2088088022191744886698175102976 \\
44 & 3738100459121645858804962689024 \\
45 & 5610148109882451164480363560960 \\
46 & 7107243167143739985776879861760 \\
47 & 6755399440834383071115485380608 \\
48 & 5629499534323800464442257309696 \\
\hline
\end{tabular}
\end{table}

\vfill\eject

\pagestyle{plain}
\def\beff{\beta_{\rm eff}}
\noindent
{\bf FIGURE CAPTIONS}
\vskip 0.5 cm
\begin{itemize}
\item [\bf Fig.~1.] The distribution of zeroes of the partition function
$\tilde{Z}(k,2,32)$ in the complex plane $(Re(k),Im(k))$.
\bigskip
\item [\bf Fig.~2.] The distribution of zeroes of the partition function
$\tilde{Z}(k,3,16)$ in the complex plane $(Re(k),Im(k))$.
\bigskip
\item [\bf Fig.~3.] The zeroes of $\tilde{Z}(k,S,T)$ closest to the real 
$k$--axis for various lattices of size $S \times T$: {\it crosses} refer
to lattices with $S=2$ and $T=2,3,4,6,8,10,16,32$, while {\it triangles}
refer to lattices with $S=3$ and $T=3,4,5,6,8,9,10,12,16$.
\bigskip
\item [\bf Fig.~4.]  The re--scaled chiral condensate,
$\tilde{Q} = - <\bar{\tilde{\psi}}\tilde{\psi}>$, as a function of $k$, for 
the two lattices $2 \times 32$ and $3 \times 16$.
\bigskip
\item [\bf Fig.~5.]  The re--scaled chiral susceptibility $\tilde{\chi}$
(defined in the text by eq.(28)) as a function of $k$, for 
the two lattices $2 \times 32$ and $3 \times 16$.
\bigskip
\item [\bf Fig.~6.]  The re--scaled third derivative $\tilde{\varphi}(k)$ 
of $ln(Z)$ with respect to $1/2k$ (defined in the text by eq.(29)), for 
the two lattices $2 \times 32$ and $3 \times 16$.
\end{itemize}

\vfill\eject
\centerline{\bf Figure 1}
\vskip 1truecm
\begin{figure}[htb]\vskip160mm
\includegraphics{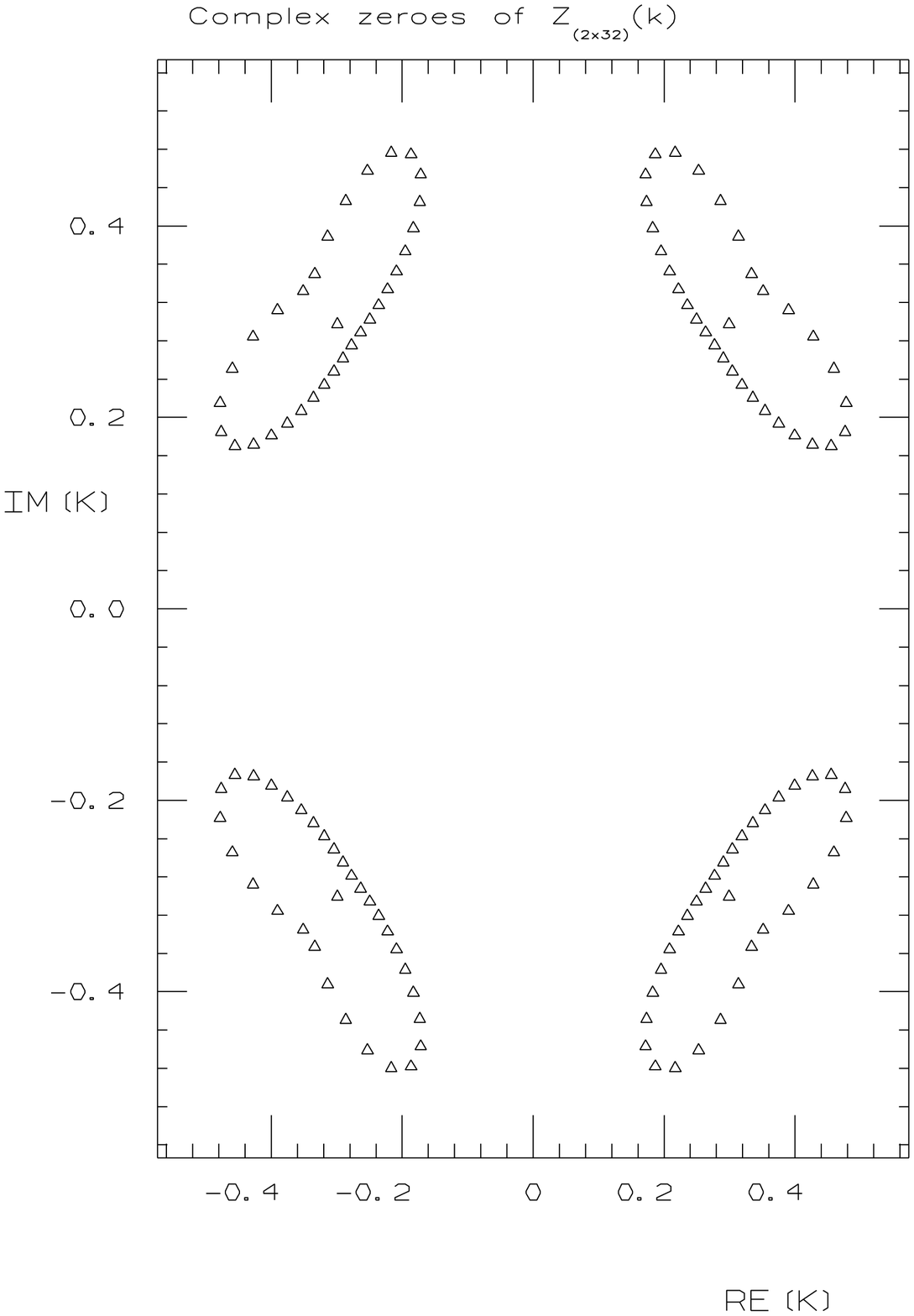}
\end{figure}

\vfill\eject
\centerline{\bf Figure 2}
\vskip 1truecm
\begin{figure}[htb]\vskip160mm
\includegraphics{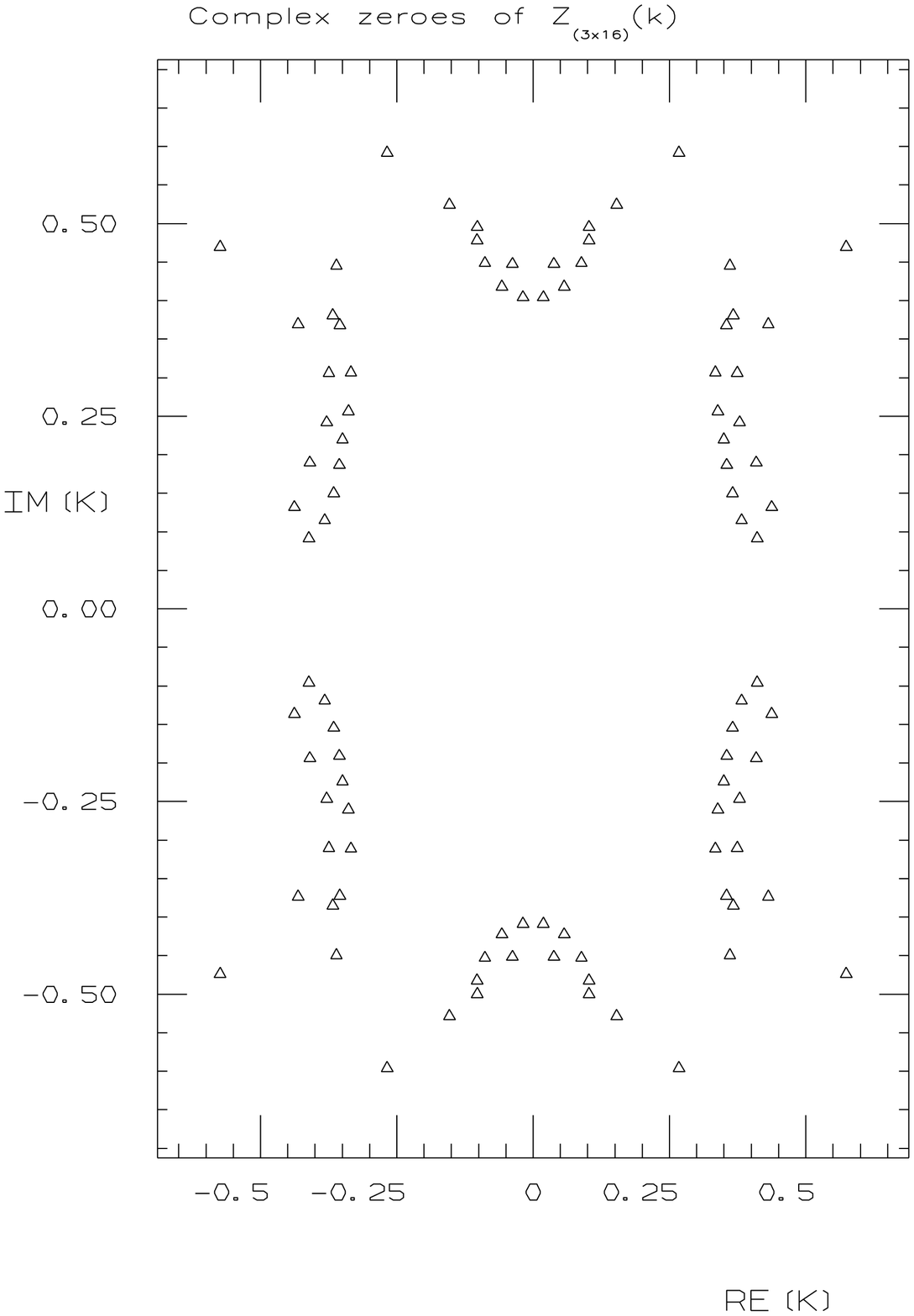}
\end{figure}

\vfill\eject
\centerline{\bf Figure 3}
\vskip 1truecm
\begin{figure}[htb]\vskip160mm
\includegraphics{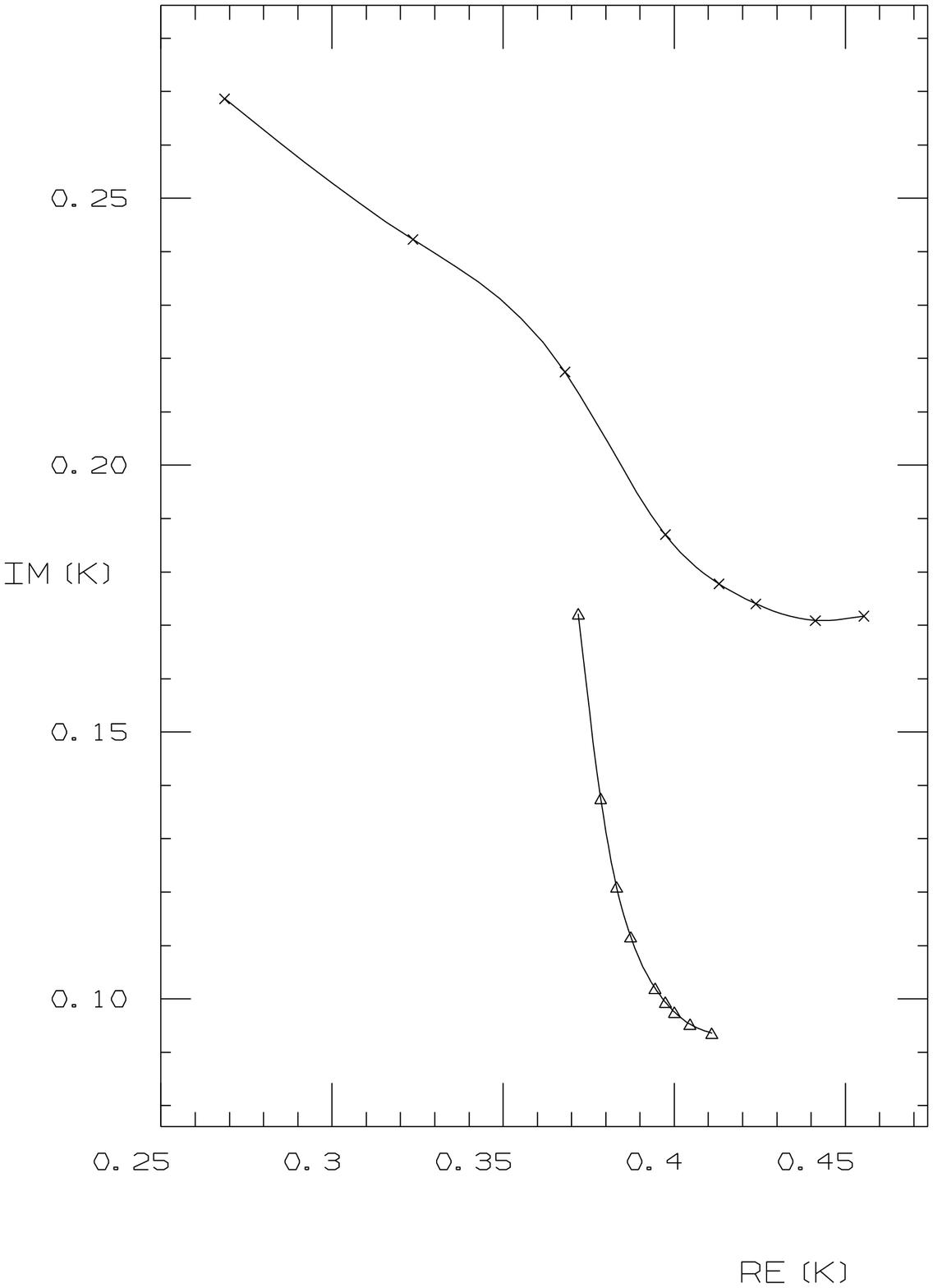}
\end{figure}

\vfill\eject
\centerline{\bf Figure 4}
\vskip 1truecm
\begin{figure}[htb]\vskip160mm
\includegraphics{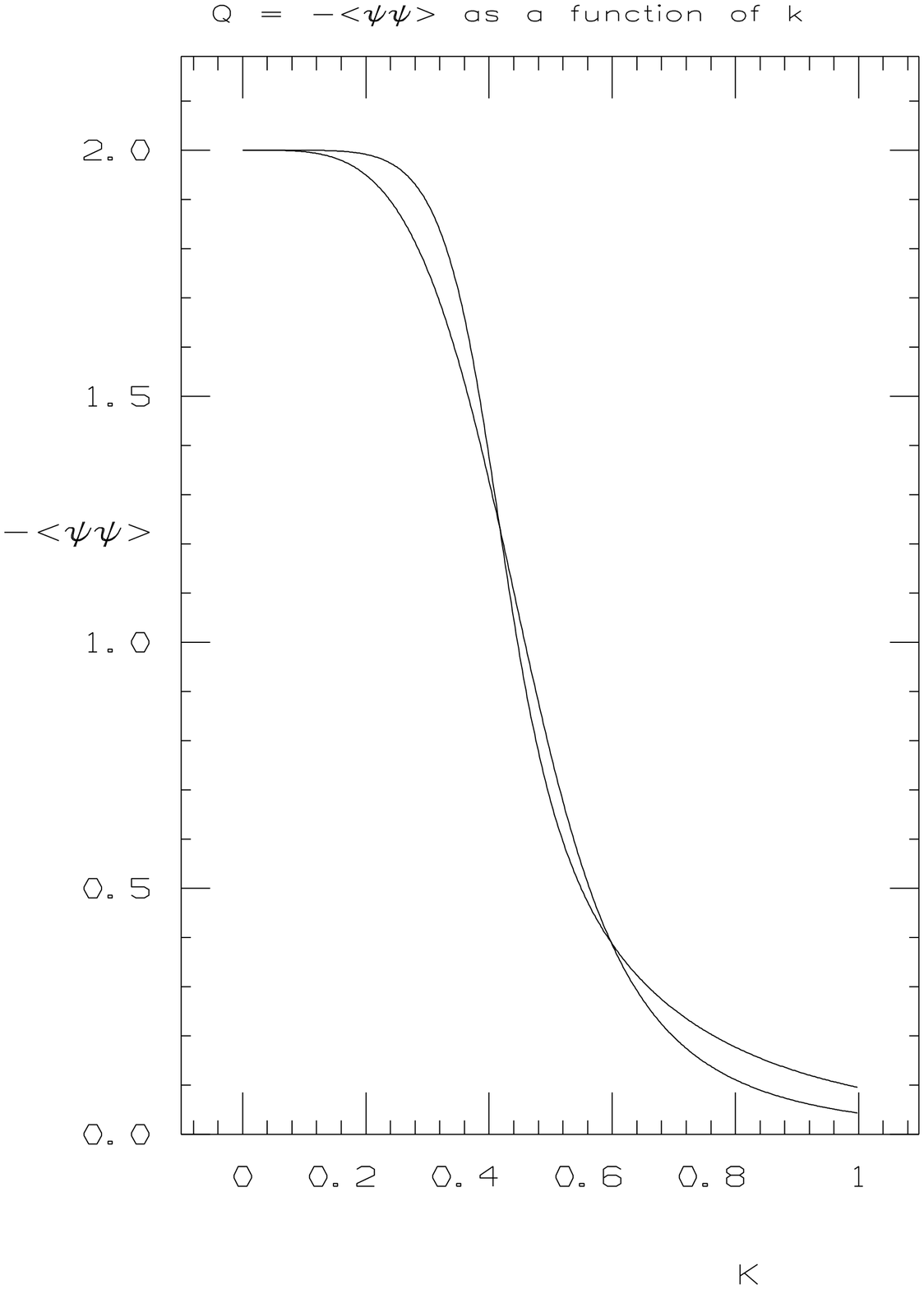}
\end{figure}

\vfill\eject
\centerline{\bf Figure 5}
\vskip 1truecm
\begin{figure}[htb]\vskip160mm
\includegraphics{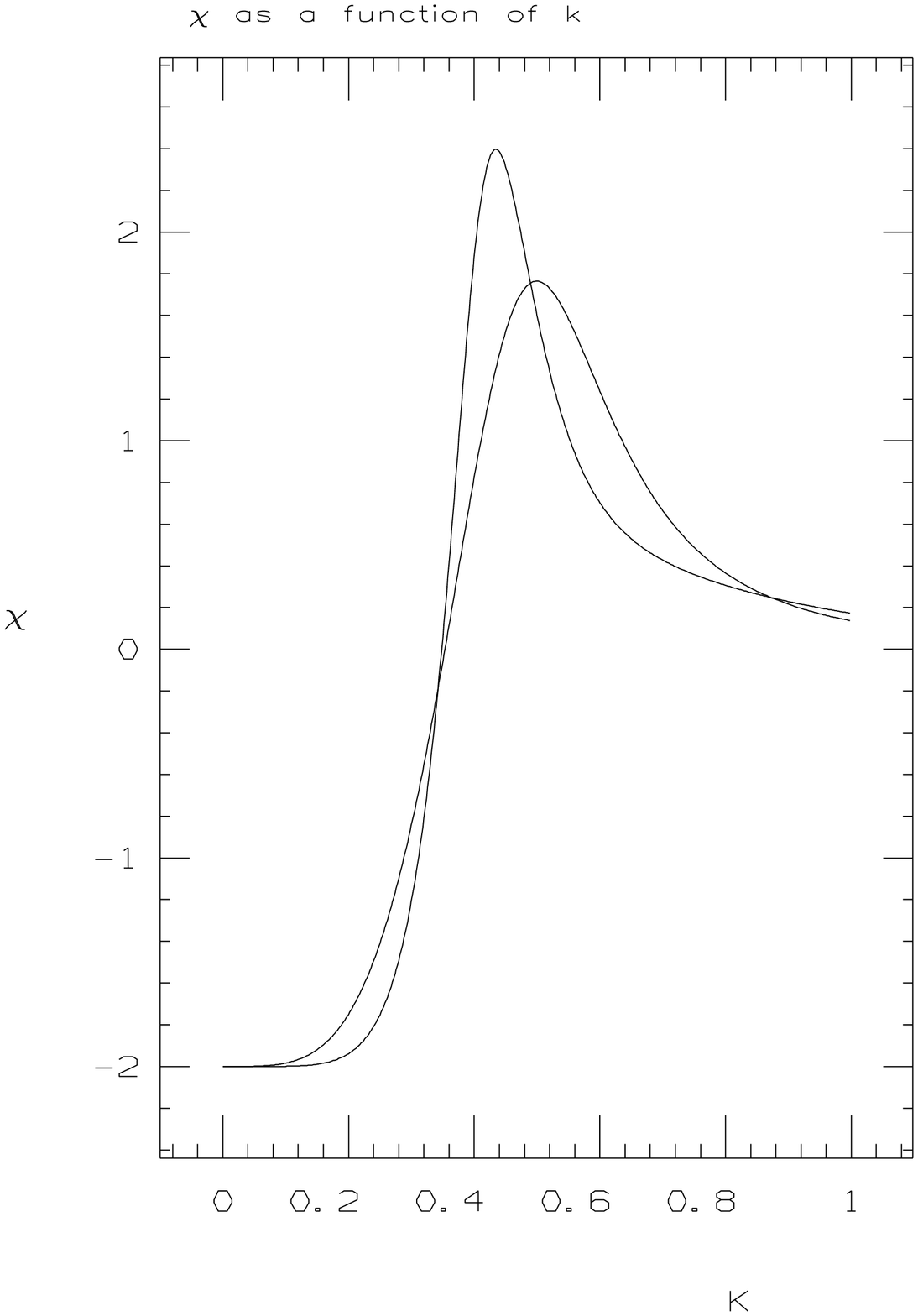}
\end{figure}

\vfill\eject
\centerline{\bf Figure 6}
\vskip 1truecm
\begin{figure}[htb]\vskip160mm
\includegraphics{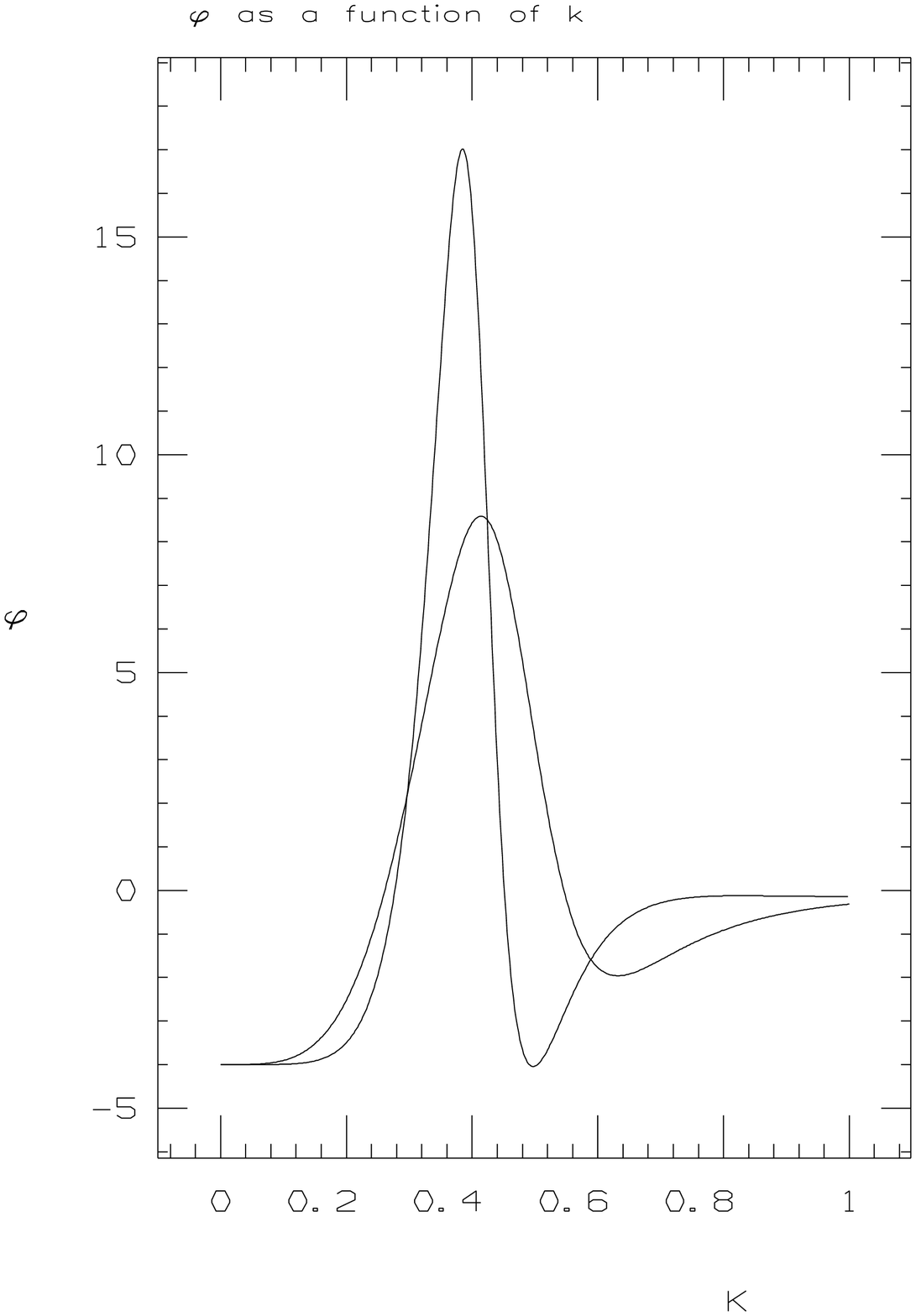}
\end{figure}

\vfill\eject
\pagestyle{empty}

\end{document}